\documentclass[12pt, preprint]{aastex}

\usepackage{graphicx}
\usepackage{natbib}
\usepackage{lscape}
\bibpunct{(}{)}{;}{a}{}{,}
\shorttitle{Near-IR $[$\ion{Se}{3}$]$ and $[$\ion{Kr}{6}$]$ Emission Lines in PNe}
\shortauthors{Sterling et al.}

\begin{document}

\title{Identification of Near-Infrared $[$\ion{Se}{3}$]$ and $[$\ion{Kr}{6}$]$ Emission Lines in Planetary Nebulae\footnote{This paper includes data obtained with the 6.5-m Magellan Telescopes at Las~Campanas Observatory, Chile.}}

\author{N.\ C.\ Sterling\altaffilmark{1}, S.\ Madonna\altaffilmark{2}, K.\ Butler\altaffilmark{3}, J.\ Garc\'ia-Rojas\altaffilmark{2}, A.\ L.\ Mashburn\altaffilmark{1}, C.\ Morisset\altaffilmark{4}, V.\ Luridiana\altaffilmark{2}, \& I.\ U.\ Roederer\altaffilmark{5, 6}}

\altaffiltext{1}{Department of Physics, University of West Georgia, 1601 Maple Street, Carrollton, GA 30118, USA; nsterlin@westga.edu, awhite15@my.westga.edu}
\altaffiltext{2}{Instituto de Astrof\'{i}sica de Canarias, E-38205 La Laguna, Tenerife, Spain; Universidad de La Laguna, Dpto. Astrof\'isica, E-38206 La Laguna, Tenerife, Spain; smadonna@iac.es, jogarcia@iac.es, vale@iaa.es}
\altaffiltext{3}{Institut f\"{u}r Astronomie und Astrophysik, Scheinerstr. 1, 81679 M\"{u}nchen, Germany; butler@usm.uni-muenchen.de}
\altaffiltext{4}{Instituto de Astronom\'ia, Universidad Nacional Autonoma de Mexico, Apdo.\ Postal 20164, 04510, Mexico; chris.morisset@gmail.com}
\altaffiltext{5}{Department of Astronomy, University of Michigan, 1085 South University Avenue, Ann Arbor, MI 48109, USA; iur@umich.edu}
\altaffiltext{6}{Joint Institute for Nuclear Astrophysics and Center for the Evolution of the Elements (JINA-CEE), USA}

\begin{abstract}

We identify $[$\ion{Se}{3}$]$~1.0994~\micron\ in the planetary nebula (PN) NGC~5315 and $[$\ion{Kr}{6}$]$~1.2330~\micron\ in three PNe, from spectra obtained with the FIRE spectrometer on the 6.5-m Baade Telescope.  Se and Kr are the two most widely-detected neutron-capture elements in astrophysical nebulae, and can be enriched by \emph{s}-process nucleosynthesis in PN progenitor stars.  The detection of $[$\ion{Se}{3}$]$~1.0994~\micron\ is particularly valuable when paired with observations of $[$\ion{Se}{4}$]$~2.2858~\micron, as it can be used to improve the accuracy of nebular Se abundance determinations, and allows Se ionization correction factor (ICF) schemes to be empirically tested for the first time.  We present new effective collision strength calculations for Se$^{2+}$ and Kr$^{5+}$, which we use to compute ionic abundances.  In NGC~5315, we find that the Se abundance computed from Se$^{3+}$/H$^+$ is lower than that determined with ICFs that incorporate Se$^{2+}$/H$^+$.  We compute new Kr ICFs that take Kr$^{5+}$/H$^+$ into account, by fitting correlations found in grids of Cloudy models between Kr ionic fractions and those of more abundant elements, and use these to derive Kr abundances in four PNe.  Observations of $[$\ion{Se}{3}$]$ and $[$\ion{Kr}{6}$]$ in a larger sample of PNe, with a range of excitation levels, are needed to rigorously test the ICF prescriptions for Se and our new Kr ICFs.

\end{abstract}


\section{INTRODUCTION} \label{intro}

Nebular spectroscopy of neutron(\emph{n})-capture elements (atomic number $Z>30$) provides a valuable tool for studying slow \emph{n}-capture nucleosynthesis (the \emph{s}-process) in low-mass stars, revealing information independent of and complementary to spectroscopy of asymptotic giant branch (AGB) stars.  In the last 15 years, this field has rapidly developed due to the discovery and detection of several \emph{n}-capture element emission lines, predominantly in the near-infrared.  \citet{dinerstein01} identified two of the most widely-detected trans-iron element lines in astrophysical nebulae, $[$\ion{Kr}{3}$]$~2.1980 and $[$\ion{Se}{4}$]$~2.2858~\micron.\footnote{Vacuum wavelengths are used throughout this paper}  More recently, \citet{sterling16} discovered emission lines of $[$\ion{Rb}{4}$]$, $[$\ion{Cd}{4}$]$, and $[$\ion{Ge}{6}$]$ in the $H$ and $K$~band spectra of two PNe.

\defcitealias{sterling15}{SPD15}

Neutron-capture elements have been detected in more than 100 PNe in the Milky Way \citep{sharpee07, sterling08, garcia-rojas15} and in nearby galaxies \citep{otsuka11, mashburn16}.  These observations have driven efforts to determine atomic data needed to accurately derive \emph{n}-capture element abundances \citep[e.g.,][]{sterling11b, sterling11c, sterling11d, sterling16}.  Detecting multiple ions of \emph{n}-capture elements leads to more accurate abundance determinations and facilitates tests of atomic data and ionization correction factor (ICF) schemes \citep[e.g.,][hereafter SPD15]{sterling15}.

In this paper, we present the first detection of $[$\ion{Se}{3}$]$~1.0994~\micron\ in an astrophysical nebula, the PN NGC~5315, and identify $[$\ion{Kr}{6}$]$~1.2330~\micron\ in four PNe.  We compute collision strengths for these ions to determine their ionic abundances.  When combined with observations of $[$\ion{Se}{4}$]$, the newly-detected $[$\ion{Se}{3}$]$ feature enables more accurate nebular Se abundance determinations, and can be used to test the Se ICF formulae of \citetalias{sterling15} for the first time.  Kr$^{5+}$ can be populated in high-ionization objects, leaving Kr$^+$ (whose sole ground configuration transition at 1.8622~\micron\ can only be detected from space) as the only significantly populated Kr ion that has not been detected in PNe.

\section{OBSERVATIONS AND DATA REDUCTION}

We observed four PNe (NGC~3918, NGC~5315, and SMP~47 and SMP~99 in the Large Magellanic Cloud or LMC) with the Folded-Port InfraRed Echellette (FIRE) spectrograph \citep{simcoe13} on the 6.5-m Baade~Telescope at Las~Campanas Observatory, Chile.  Each object was observed in echelle mode with a 0\farcs75$\times$7\arcsec\ slit, for a spectral resolution $R=4800$ over the wavelength range 0.83--2.45~\micron.  Details of the observations and analysis of LMC~SMP~47 and SMP~99 are discussed in \citet{mashburn16}.  NGC~3918 was observed on 2013-01-22 and NGC~5315 on 2013-08-13, using on-off sequences for sky subtraction, for total on-source integration times of 1100 and 600~s, respectively.  We centered the slit on the central portion of each nebula, at PA~0$^o$ for NGC~3918 and 90$^o$ for NGC~5315.  For each PN, an A0V standard with similar airmass was observed for telluric corrections and relative flux calibrations, and Th-Ar lamp spectra were used for wavelength calibration.  The data were reduced using the FIREHOSE~IDL reduction pipeline.\footnote{Available at http://web.mit.edu/\~{}rsimcoe/www/FIRE/}  A detailed analysis of the FIRE spectrum of NGC~5315 is given in \citet{madonna17}, and that for NGC~3918 will be described in a future paper.


\section{LINE IDENTIFICATIONS}

\subsection{$[$\ion{Se}{3}$]$~1.0994~\micron}

\begin{figure}[ht!]
\epsscale{0.7}
\plotone{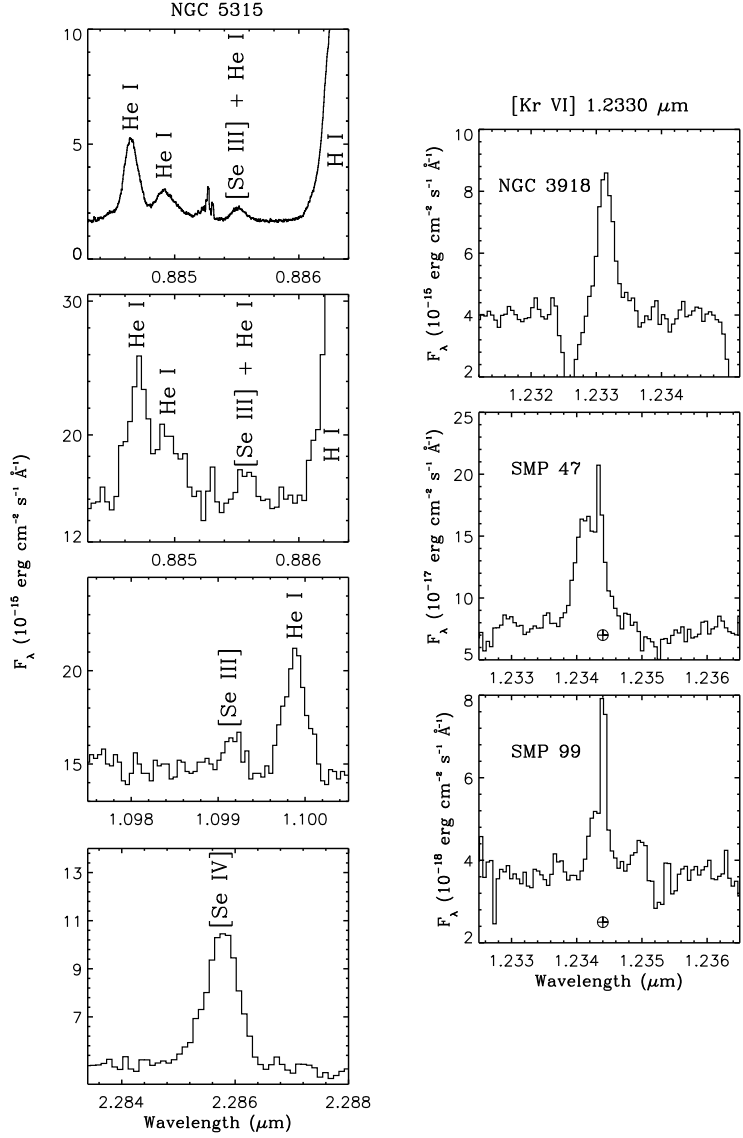}
\caption{\textit{Left panels: }$[$\ion{Se}{3}$]$ and $[$\ion{Se}{4}$]$ detections in the UVES \citep[top panel; ][]{madonna17} and FIRE spectra of NGC~5315.  \textit{Right panels: }$[$\ion{Kr}{6}$]$~1.2330~\micron\ detections in NGC~3918 and the LMC PNe SMP~47 and SMP~99.  Telluric lines (removed by Gaussian fits in the flux measurements) are indicated by $\Earth$ symbols.}
\label{detections}
\end{figure}

With an ionization potential range of 21.2--31.7~eV, Se$^{2+}$ can be significantly populated in low- to moderate-excitation PNe such as NGC~5315.  Two $[$\ion{Se}{3}$]$ lines fall in the spectral range of FIRE, 0.8858 and 1.0994~\micron, where we use the level energies of \citet{tauheed12}.  These correspond to the $^{1}$D$_2$--$^3$P$_1$ and $^{1}$D$_2$--$^3$P$_2$ ground configuration transitions, respectively.  We detect features at both wavelengths in NGC~5315, as well as $[$\ion{Se}{4}$]$~2.2858~\micron\ (Figure~\ref{detections}, Table~\ref{ionicf}).  The 0.8858~\micron\ line is also detected in the UVES spectrum presented by \citet{madonna17}, at higher signal-to-noise than in the FIRE data.

\begin{landscape}
\begin{deluxetable}{llcccccccc}
\tablecolumns{9}
\tablewidth{0pc} 
\tabletypesize{\scriptsize}
\tablecaption{Se and Kr Intensities and Abundances} 
\tablehead{
\colhead{PN} & \colhead{} & \colhead{} & \colhead{$I(\lambda)$/$I$(\ion{H}{1})\tablenotemark{a}} & \colhead{$I(\lambda)$/$I$(\ion{H}{1})$_{\rm corr}$\tablenotemark{b}} & \colhead{X$^{+i}$/H$^+$} & \colhead{} & \colhead{$12+$} & \colhead{$12+$}\\
\colhead{Name} & \colhead{Ion} & \colhead{$\lambda$ (\micron)} & \colhead{$\times 100$}  & \colhead{$\times 100$}& \colhead{$\times 10^{-9}$} & \colhead{ICF} & \colhead{Log(X/H)} & \colhead{Log(X/H)$_{\mathrm{Lit}}$}}
\startdata
NGC 5315   & $[$\ion{Se}{3}$]$ & 0.8858\tablenotemark{c} & 0.18$\pm$0.05 & 0.10$\pm$0.05 & 1.3$\pm$0.7   & 3.6$\pm$2.8 & 3.67$\pm$0.45 & \nodata \\
           & $[$\ion{Se}{3}$]$ & 0.8858 & 0.20$\pm$0.07 & 0.11$\pm$0.07 & 1.5$\pm$0.8   &             & 3.72$\pm$0.45 &               \\
           & $[$\ion{Se}{3}$]$ & 1.0994 & 0.21$\pm$0.07 & \nodata       & 2.1$\pm$0.8   &             & 3.89$\pm$0.27 &               \\
           & $[$\ion{Se}{4}$]$ & 2.2858 & 2.95$\pm$0.32 & \nodata       & 0.80$\pm$0.08 & 1.7$\pm$0.5 & 3.12$\pm$0.13 &               \\
           &                   &        &               &               &               & 1.3$\pm$0.5\tablenotemark{d} & 3.70$\pm$0.21\tablenotemark{d} &  \\
NGC 3918   & $[$\ion{Kr}{6}$]$ & 1.2330 & 0.44$\pm$0.06 & \nodata       & 0.18$\pm$0.03 & 5.1$\pm$0.1 & 2.95$\pm$0.08 & 3.97$\pm$0.12 \\
           &                   &        &               &               &               & 1.2$\pm$0.1\tablenotemark{e} & 3.90$\pm$0.07\tablenotemark{e} &  \\
LMC SMP 47 & $[$\ion{Kr}{6}$]$ & 1.2330 & 0.46$\pm$0.06 & 0.22$\pm$0.11 & 0.08$\pm$0.06 & 7.8$\pm$3.2 & 2.82$\pm$0.39 & 3.75$\pm$0.20  \\
LMC SMP 99 & $[$\ion{Kr}{6}$]$ & 1.2330 & 0.21$\pm$0.04 & \nodata & 0.11$\pm$0.03 & 36$\pm$21 & 3.59$\pm$0.34 &  3.78$\pm$0.21 \\
NGC 7027\tablenotemark{f}   & $[$\ion{Kr}{6}$]$ & 1.2330 & 1.1$\pm$0.3   & 1.0$\pm$0.3   & 0.53$\pm$0.16 & 5.0$\pm$2.6 & 3.42$\pm$0.24 & 4.01$\pm$0.10 \\
           &                   &        &               &               &               & 1.2$\pm$0.1\tablenotemark{e} & 4.10$\pm$0.04\tablenotemark{e} &  \\
\tableline
\enddata
\label{ionicf}
\tablecomments{Intensities and ionic and elemental abundances are given for each Se line in NGC~5315, using the ICF schemes of \citetalias{sterling15}: Equation~7 for $[$\ion{Se}{3}$]$ lines and Equation~8 for $[$\ion{Se}{4}$]$.  Elemental Kr abundances were derived from Kr$^{5+}$/H$^+$ using Equation~\ref{icf_kr6} (\S\ref{kr6_sect}).  Kr abundances from the literature (the final column) are from \citet{garcia-rojas15} for NGC~3918, \citet{mashburn16} for the LMC PNe, and \citetalias{sterling15} for NGC~7027.}
\tablenotetext{a}{We use \ion{H}{1}~Pa$\delta$ as the reference line for $[$\ion{Se}{3}$]$, Br$\gamma$ for $[$\ion{Se}{4}$]$, and Pa$\beta$ for $[$\ion{Kr}{6}$]$.}
\tablenotetext{b}{$[$\ion{Se}{3}$]$~0.8858~\micron\ intensity corrected for contamination by \ion{He}{1}~0.8857~\micron, and $[$\ion{Kr}{6}$]$~1.2330~\micron\ intensities corrected (if necessary) for H$_2$~3-1~S(1)~1.2330~\micron.  The error bars account for uncertainties in the deblending procedure.}
\tablenotetext{c}{Intensity from UVES data \citep{madonna17}.}
\tablenotetext{d}{Elemental Se abundance computed from (Se$^{2+}$~+~Se$^{3+}$)/H$^+$, using Equation~9 of \citetalias{sterling15} and Se$^{2+}$/H$^+$ from the 1.0994~\micron\ line.}
\tablenotetext{e}{Elemental Kr abundance computed from Equation~\ref{icf_kr3456} (\S\ref{kr6_sect}), using the Kr$^{2+}$--Kr$^{4+}$ ionic abundances from \citet{garcia-rojas15} for NGC 3918 and \citet{sharpee07} for NGC 7027.}
\tablenotetext{f}{Intensity from \citet{kelly95}, with an assumed uncertainty of 30\%.}
\end{deluxetable}
\end{landscape}

We identify the weak feature at 1.0992~\micron\ as $[$\ion{Se}{3}$]$~1.0994~\micron.  The radial velocity is --56~km~s$^{-1}$, which agrees well with lines in the UVES spectrum but is slightly larger than found for other lines in the FIRE data \citep[typically --20 to --40~km~s$^{-1}$;][]{madonna17}.

Using the Atomic Line List\footnote{http://www.pa.uky.edu/$\sim$peter/newpage/}, we searched for other possible identifications within $\pm$10~\AA\ of the observed wavelength.  We considered forbidden lines with excitation energies up to 10~eV and permitted lines for elements in the first three rows of the Periodic Table.  No multiplet members or lines from the same upper level were found for any of the potential alternative identifications, except for \ion{He}{1}~1.1000~\micron\ which is easily resolved from the 1.0992~\micron\ feature (Figure~\ref{detections}).  The strongest molecular transition with a similar wavelength is H$_2$~2--0~S(4) 1.0998~\micron.  However, NGC~5315 does not exhibit emission from vibrationally-excited H$_2$ in its FIRE spectrum \citep{madonna17}, making this identification unlikely.

Therefore, $[$\ion{Se}{3}$]$ is the most probable identification for the 1.0992~\micron\ feature.  Because this line shares the same upper level as $[$\ion{Se}{3}$]$~0.8858~\micron, the relative intensities of the two lines should equal the ratio of the transition probabilities (\S\ref{calcs}) multiplied by the relative energies of the lines: $I$(1.0994)/$I$(0.8858)~=~1.3.  A direct comparison is complicated because the 0.8858~\micron\ feature is contaminated by \ion{He}{1}.  

\begin{figure}[ht!]
\epsscale{0.9}
\plotone{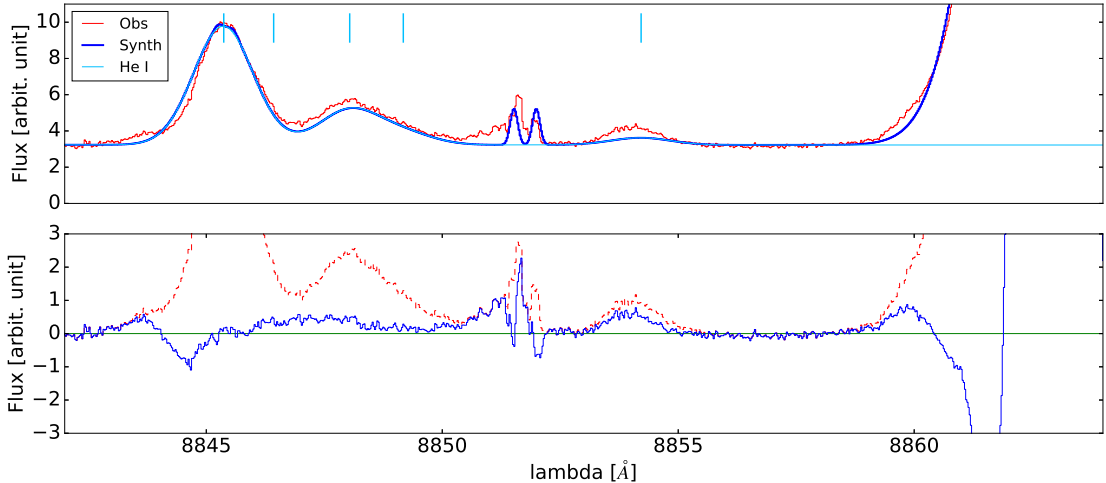}
\caption{PySSN fit to the NGC~5315 spectrum near $[$\ion{Se}{3}$]$, with flux in arbitrary units and wavelengths in air.  The upper panel compares the synthetic \ion{H}{1} and \ion{He}{1} spectrum (blue line) to the observed UVES spectrum of \citet{madonna17}.  In the lower panel, the residuals of the fit (blue line) are plotted against the observed spectrum (red dashed line).  A feature near 0.8854~\micron\ is visible in the residuals, corresponding to $[$\ion{Se}{3}$]$.}
\label{8858_fit}
\end{figure}

To determine the $[$\ion{Se}{3}$]$ contribution to the 0.8858~$\mu$m feature, we utilized the PySSN spectrum synthesis code (v0.2.70) to remove the \ion{He}{1} contribution.  PySSN is a python version of X-SSN \citep{pequignot12}, which uses a database of transitions including relative intensities of multiplet members to generate a synthetic spectrum.  It takes into account the different line profiles for various ions and emission processes, interstellar extinction, and instrumental response.  In Figure~\ref{8858_fit}, we plot the PySSN synthetic \ion{H}{1} and \ion{He}{1} spectrum against the UVES data of \citet{madonna17} in the region near 0.8858~$\mu$m.\footnote{The PySSN fit is in air wavelengths, since the v0.2.70 transition database utilizes only air wavelengths in the optical and $I$~band.}  The \ion{He}{1} contribution ($\sim$45\% of the measured flux) is removed in the bottom panel, and the residuals show a clear feature corresponding to $[$\ion{Se}{3}$]$.

Correcting for the \ion{He}{1} contribution results in $I$(1.0994)/$I$(0.8858)~=~2.0$\pm$1.3.  This is somewhat larger than predicted, likely due to the uncertain correction for \ion{He}{1} to the weak 0.8858~\micron\ line.  Nevertheless, the observed and expected ratios agree within the uncertainties, supporting our identification.

\subsection{$[$\ion{Kr}{6}$]$~1.2330~\micron}

In the FIRE spectra of NGC~3918 and the LMC PNe SMP~47 and SMP~99, we detect a line at 1.2330~\micron\ (rest wavelength) that we identify as $[$\ion{Kr}{6}$]$~1.2330~\micron.  This is the only collisionally-excited transition of Kr$^{5+}$, whose ground configuration has a single $^2$P term.   Kr$^{5+}$ has an ionization potential range of 64.7--78.5~eV, and thus is detectable in high-ionization nebulae such as these three objects.  

\begin{figure}[t!]
\epsscale{0.9}
\plotone{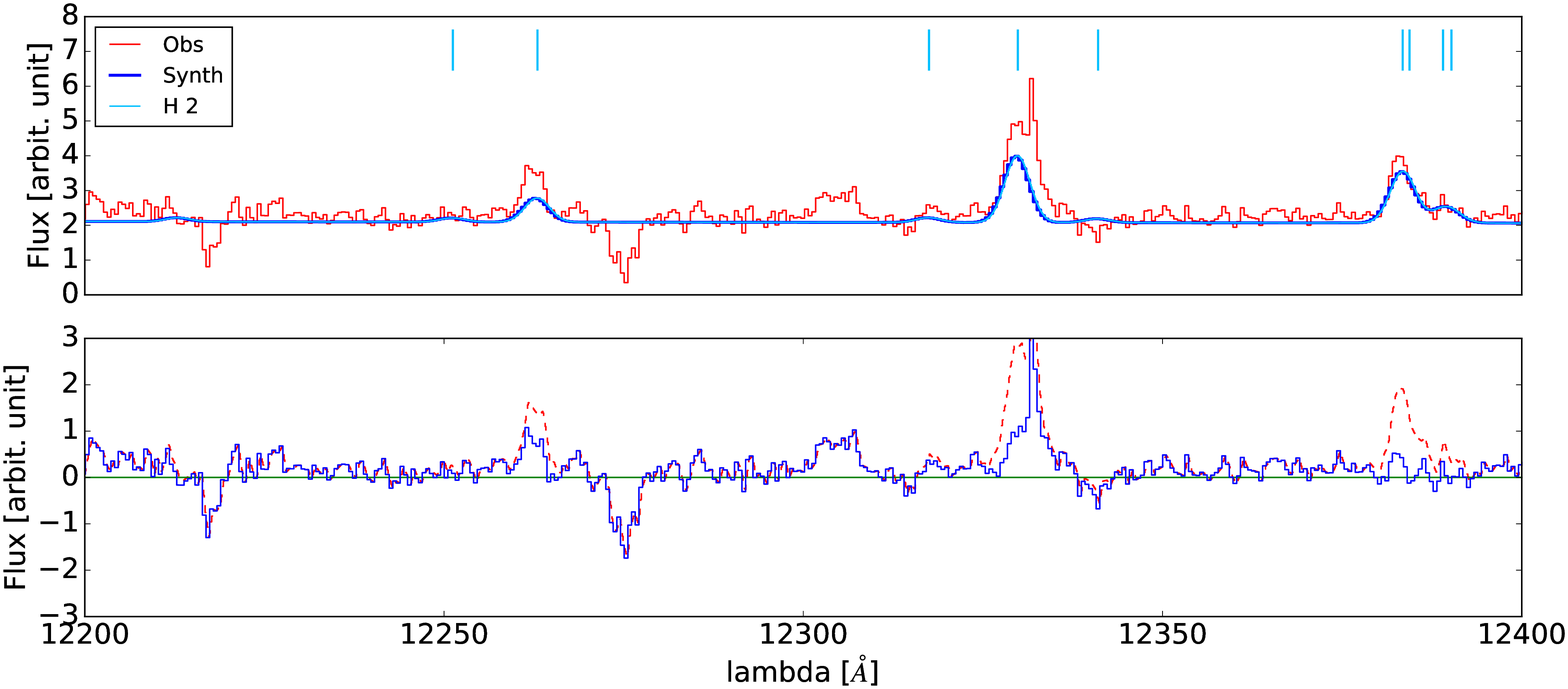}
\caption{PySSN fit to the SMP~47 spectrum near $[$\ion{Kr}{6}$]$~1.2330~\micron, with flux in arbitrary units and vacuum wavelengths corrected for the systemic velocity of the nebula.  The upper panel compares the synthetic H$_2$ spectrum \citep[blue line, with wavelengths from][]{abgrall00} to the observed spectrum, with line strengths scaled to H$_2$~3-1~S(2)~1.2076~\micron\ using the relative intensities from model~14 of \citet{bvd87}.  Note that $[$\ion{Fe}{2}$]$~1.2304~\micron\ is not included in the synthetic fit.  In the lower panel, the residuals of the fit (blue line) show an emission feature at 1.2330~\micron, which we identify as $[$\ion{Kr}{6}$]$.}  
\label{12330_fit}
\end{figure}

There are several possible identifications for the 1.2330~\micron\ feature, most notably H$_2$~3-1~S(1)~1.2330, $[$\ion{Fe}{6}$]$~1.2330, and \ion{N}{1}~1.2329~\micron\ (see below).  For each of the observed PNe, we did not detect multiplet members or lines from the same upper level for any atomic transitions within 10~\AA\ of the observed wavelength.  SMP~47 exhibits vibrationally-excited H$_2$ emission consistent with fluorescent excitation in a moderate density gas \citep{mashburn16}, and H$_2$ likely contributes to the 1.2330~\micron\ feature in this PN.  According to model~14 of \citet{bvd87}, the H$_2$ 3-1~S(1) line is 1.9 times stronger than 3-1~S(2)~1.2076~\micron, and accounts for approximately 50\% of the 1.2330~\micron\ flux in SMP~47.  In Figure~\ref{12330_fit} we show a PySSN fit of the H$_2$ contribution to the 1.2330~\micron\ feature in SMP~47, set to 1.9$F$(H$_2$~3-1~S(2)).  The bottom panel shows the residual emission due to $[$\ion{Kr}{6}$]$.  NGC~3918 and SMP~99 do not exhibit vibrationally-excited H$_2$ emission, and no correction is needed for those PNe.

A feature at 1.2330~\micron\ has been previously detected in other objects, but is likely due to $[$\ion{Kr}{6}$]$ only in the PN NGC~7027 \citep[][Table~\ref{ionicf}]{rudy92, kelly95}.  In low-excitation objects such as Hb~12, BD+30$^o$3639, M~1-11, and the Orion nebula \citep{kelly95, hora99, otsuka13a}, the intensity of 1.2330~\micron\ relative to H$_2$ lines indicates that it can be identified as H$_2$~3-1~S(1).  The detection of $[$\ion{Fe}{6}$]$ lines in the $J$~band spectra of classical novae \citep[e.g.,][]{lynch01, rudy02}, including multiplet members of $[$\ion{Fe}{6}$]$~1.2330~\micron, suggests that the 1.2330~\micron\ line in these objects can be attributed to $[$\ion{Fe}{6}$]$.  In the low-excitation PN SMC~Lin~49 \citep{otsuka16}, the line is likely due to \ion{N}{1}~1.2329~\micron, based on the detection of both multiplet members.  These examples illustrate the need for caution in identifying the 1.2330~\micron\ feature as $[$\ion{Kr}{6}$]$.

\section{COLLISION STRENGTH CALCULATIONS} \label{calcs}

The collisional data for Se$^{2+}$ and Kr$^{5+}$ were calculated using the R-matrix method \citep{berrington95}.  Relativistic effects (mass, velocity, and Darwin terms) were included by using the intermediate-coupling frame transformation method \citep{griffin98}.  The radial scaling parameters for the target ion and the scattering target were obtained in the configuration interaction approximation with AUTOSTRUCTURE \citep{badnell11}.
  
Details and results of the calculations are given in Tables~\ref{energies} and \ref{atomic_data}.  We utilized kappa-averaged relativistic wavefunctions, which give markedly better energy levels and transition probabilities for these ions although they have little effect on the collisional data.  The lambda parameters scale a Thomas-Fermi-Dirac-Almadi potential for each of the electron orbitals, and were determined using a two-step process.  First, the $1s$ through $4p$ scaling parameters were determined by minimizing the energies of LS terms in the ground configuration of each ion, and then these scaling parameters were held fixed when computing those for other orbitals on the first 23 (Se$^{2+}$) and 24 (Kr$^{5+}$) target term energies.  The calculated energies are in excellent agreement (9\% or better) with observed values from NIST \citep{nist}.  Our Se$^{2+}$ transition probabilities agree with those of \citet{biemont86a} to within 15\% or better, with the exception of the weak E2 transition $^1$D$_2$--$^3$P$_0$ (60\%), and similar agreement is found for Kr$^{5+}$ with \citet{biemont87}.

\begin{deluxetable}{ccccc}
\tablecolumns{5}
\tablewidth{0pc} 
\tabletypesize{\scriptsize}
\tablecaption{Comparison of Calculated and Experimental Energies}
\tablehead{
\colhead{Config.} & \colhead{Level} & \colhead{Energy (NIST, cm$^{-1}$)} & \colhead{Energy (cm$^{-1}$)} & \colhead{Diff.\ (\%)}}
\startdata
\multicolumn{5}{c}{Se$^{2+}$\tablenotemark{a}} \\
\cline{1-5}
$4s^24p^2$ & $^3$P$_0$ & 0      & 0      &       \\
           & $^3$P$_1$ & 1741   & 1700   & --2.4 \\
           & $^3$P$_2$ & 3937   & 3914   & --0.6 \\
           & $^1$D$_2$ & 13032  & 14211  & 9.0   \\
           & $^1$S$_0$ & 28430  & 29985  & 5.5   \\
$4s4p^3$   & $^5$S$_2$ &        & 61473  &       \\
           & $^3$D$_1$ & 91091  & 89796  & --1.4 \\
           & $^3$D$_2$ & 92723  & 89868  & --3.1 \\
           & $^3$D$_3$ & 96548  & 90346  & --6.4 \\
           & $^3$P$_0$ & 106482 & 104358 & --2.0 \\
           & $^3$P$_2$ & 106515 & 104543 & --1.9 \\
           & $^3$P$_1$ & 106591 & 104579 & --1.9 \\
\cline{1-5}
\multicolumn{5}{c}{Kr$^{5+}$\tablenotemark{b}} \\
\cline{1-5}
$4s^24p$ & $^2$P$_{1/2}$ & 0       & 0       & \\
         & $^2$P$_{3/2}$ & 8110    & 7774    & --4.3 \\
$4s4p^2$ & $^4$P$_{1/2}$ & 107836  & 104288  & --3.4 \\
         & $^4$P$_{3/2}$ & 111193  & 107442  & --3.5 \\
         & $^4$P$_{3/2}$ & 115479  & 111589  & --3.5 \\
         & $^2$D$_{5/2}$ & 141672  & 140206  & --1.0 \\
         & $^2$D$_{3/2}$ & 142727  & 141068  & --1.2 \\
         & $^2$S$_{5/2}$ & 170084  & 172042  & 1.1 \\
         & $^2$P$_{1/2}$ & 180339  & 184060  & 2.0 \\
         & $^2$P$_{1/2}$ & 183817  & 187921  & 2.1 \\
\tableline
\enddata
\label{energies}
\tablenotetext{a}{The configuration expansion for Se$^{2+}$ is: 4$s^2$4$p^2$, 4$s$4$p^3$, 4$s^24p4d$, 4$s^2$4$d^2$, 4$s4p^24d$, $4s4p4d^2$, 4$p^4$, 4$p^34d$, 4$p^24d^2$, 4$s^24p4f$, 4$s^24p5s$, 4$s^24p5p$, 4$s4p^24f$, 4$s4p^25s$, 4$s4p^25p$, 4$p^34f$, 4$p^35s$, 4$p^34f$, 4$p^35p$.  The scaling parameters used are 1.42629 ($1s$), 1.14035 ($2s$), 1.08559 ($2p$), 1.03542 ($3s$), 1.01549 ($3p$), 0.99471 ($3d$), 0.96839 ($4s$), 0.97685 ($4p$), 1.01920 ($4d$), 1.41658 ($4f$), 0.98618 ($5s$), and 0.98063 ($5p$).}
\tablenotetext{b}{The configuration expansion for Kr$^{5+}$ includes all 46 one and two electron promotions from the ground configuration $4s^24p$ into the $4p$, $4d$, $4f$, $5s$, $5p$, and $5d$ orbitals.  The scaling parameters, listed in the same order as those for Se$^{2+}$, are 1.42203, 1.13758, 1.08300, 1.03631, 1.01532, 0.99812, 0.98673, 0.98961, 1.00041, 1.07879, 1.01934, 1.01568, and 1.01620 ($5d$).}
\end{deluxetable}

\begin{landscape}
\begin{deluxetable}{lcccccccccccc}
\tablecolumns{13}
\tablewidth{0pc} 
\tabletypesize{\scriptsize}
\tablecaption{Se$^{2+}$ and Kr$^{5+}$ Transition Probabilities and Effective Collision Strengths} 
\tablehead{
\colhead{} & \colhead{$A_{ij}$} & \multicolumn{11}{c}{Effective Collision Strength $\Upsilon(T)$} \\ \cline{3-13}
\colhead{Trans.} & \colhead{(s$^{-1}$)} & \colhead{2000 K} & \colhead{4000 K} & \colhead{6000 K} & \colhead{8000 K} & \colhead{10000 K} & \colhead{12000 K} & \colhead{14000 K} & \colhead{16000 K} & \colhead{18000 K} & \colhead{20000 K} & \colhead{50000 K}}
\startdata
\multicolumn{13}{c}{Se$^{2+}$} \\
\cline{1-13}
$^3$P$_1$--$^3$P$_0$ & 8.708E-02 & 1.83     & 1.80     & 1.77     & 1.76     & 1.76     & 1.77     & 1.78     & 1.78     & 1.79     & 1.79     & 1.76    \\
$^3$P$_2$--$^3$P$_0$ & 1.693E-04 & 1.14     & 1.10     & 1.10     & 1.11     & 1.13     & 1.15     & 1.18     & 1.21     & 1.24     & 1.27     & 1.46 \\
$^3$P$_2$--$^3$P$_1$ & 1.418E-01 & 4.60     & 4.46     & 4.49     & 4.55     & 4.62     & 4.70     & 4.78     & 4.87     & 4.94     & 5.02     & 5.47  \\
$^1$D$_2$--$^3$P$_0$ & 3.821E-04 & 6.48E-01 & 6.54E-01 & 6.46E-01 & 6.44E-01 & 6.48E-01 & 6.57E-01 & 6.68E-01 & 6.79E-01 & 6.90E-01 & 7.01E-01 & 7.67E-01 \\
$^1$D$_2$--$^3$P$_1$ & 8.076E-01 & 2.04     & 2.08     & 2.06     & 2.07     & 2.08     & 2.11     & 2.15     & 2.18     & 2.22     & 2.25     & 2.43 \\
$^1$D$_2$--$^3$P$_2$ & 1.310     & 3.85     & 3.92     & 3.92     & 3.93     & 3.96     & 4.00     & 4.04     & 4.09     & 4.13     & 4.18     & 4.34 \\
$^1$S$_0$--$^3$P$_0$ & \nodata   & 1.43E-01 & 1.37E-01 & 1.39E-01 & 1.45E-01 & 1.51E-01 & 1.57E-01 & 1.62E-01 & 1.66E-01 & 1.70E-01 & 1.73E-01 & 1.78E-01 \\
$^1$S$_0$--$^3$P$_1$ & 1.756E+01 & 4.19E-01 & 4.01E-01 & 4.03E-01 & 4.18E-01 & 4.36E-01 & 4.52E-01 & 4.65E-01 & 4.75E-01 & 4.83E-01 & 4.89E-01 & 4.81E-01 \\
$^1$S$_0$--$^3$P$_2$ & 5.108E-01 & 7.21E-01 & 6.79E-01 & 6.76E-01 & 6.97E-01 & 7.24E-01 & 7.49E-01 & 7.71E-01 & 7.88E-01 & 8.01E-01 & 8.11E-01 & 7.92E-01 \\
$^1$S$_0$--$^1$D$_2$ & 3.380     & 2.16     & 1.85     & 1.75     & 1.74     & 1.77     & 1.82     & 1.88     & 1.94     & 1.99     & 2.05     & 2.42 \\
\cline{1-13}
\multicolumn{13}{c}{Kr$^{5+}$} \\
\cline{1-13}
$^2$P$_{3/2}$--$^2$P$_{1/2}$ & 4.229 & 1.11E+01 & 1.22E+01 & 1.25E+01 & 1.25E+01 & 1.23E+01 & 1.21E+01 & 1.19E+01 & 1.17E+01 & 1.15E+01 & 1.13E+01 & 9.71 \\
\tableline
\enddata
\label{atomic_data}
\tablecomments{The full collision strength results and radiative data in the adf04 format of the ADAS\footnote{http://www.adas.ac.uk/man/appxa-04.pdf} project are available as supplementary data files to this article.}
\end{deluxetable}
\end{landscape}
  
The scattering calculations for Se$^{2+}$ included 23 target terms, leading to a 41-level calculation once relativistic effects were included.  The corresponding numbers for Kr$^{5+}$ are 24 and 52.  The calculated term and level energies were replaced by NIST values when available.  We described the scattered electrons with 44 and 39 continuum orbitals for the respective ions.  The collision strengths were calculated with the UK-APAP codes\footnote{http://amdpp.phys.strath.ac.uk/UK\_RmaX/codes.html} \citep{badnell99}, using fixed energy grids with $10^{-5}z^2$~Ryd steps, where $z$ is the ion charge, to resolve resonance structures at low energies, and a coarser mesh of $0.01z^2$~Ryd at higher energies where the collision strengths are smooth.  The R-matrix calculations included exchange up to $L=12$ (corresponding to $J=9.5$ and 10 for Se$^{2+}$ and Kr$^{5+}$), and were ``topped up'' with non-exchange data up to $J=37.5$ and 38.   The sum over $J$ was completed by extrapolation procedures involving the Born approximation  and geometric sums \citep[e.g.,][]{witthoeft06}.  The resulting total collision strengths were convolved with a Maxwellian distribution for a number of temperatures to produce the effective collision strengths $\Upsilon(T)$ in Table~\ref{atomic_data}.  Downward collisional rate coefficients $q_{ji}$ from level $j$ to level $i$ can be computed via
\begin{displaymath}
     q_{ji}=\frac{8.631\times10^{-6}}{T^{1/2}g_j}\Upsilon(T),
\end{displaymath}
where $g_j$ is the statistical weight of the upper state.  Based on the agreement of the energy levels and the transition probabilities, we estimate that the error bars on the collision strengths are no larger than $\sim$30\%.

\section{RESULTS AND DISCUSSION}

We computed ionic Se abundances in NGC 5315 and Kr$^{5+}$/H$^+$ in the other PNe (Table~\ref{ionicf}), including NGC~7027 using the data of \citet{kelly95}.  The abundances were derived with PyNeb \citep{luridiana15}, using the transition probabilities and effective collision strengths from Table~\ref{atomic_data} and Se$^{3+}$ atomic data from \citet{biemont87} and K.\ Butler (2007, private communication).  We adopt electron temperatures and densities from \citet{madonna17} for NGC~5315, \citet{ld06} for the LMC PNe, \citet{garcia-rojas15} for NGC~3918, and \citet{sharpee07} for NGC~7027.  The error bars for the ionic abundances account for uncertainties in line fluxes, temperatures, and densities, propagated via Monte Carlo simulations.

\subsection{The Se Abundance in NGC~5315}

Se$^{2+}$ is the only Se ion other than Se$^{3+}$ that has been detected in PNe.  The detection of the uncontaminated $[$\ion{Se}{3}$]$~1.0994~\micron\ line (0.8858~\micron\ is often blended with a \ion{He}{1} line), allows nebular Se abundances to be determined more accurately than previously possible.  This feature can also be used to compute Se abundances in low-ionization nebulae in which $[$\ion{Se}{4}$]$ is not detected.  In addition, this detection enables the ICF formulae of \citetalias{sterling15} to be empirically tested for the first time.  Such tests are critical for verifying the accuracy of atomic data governing the ionization balance, as illustrated for the case of Kr by \citetalias{sterling15}.  The Se ICFs from \citetalias{sterling15} are:
\begin{equation}
\eqnum{7}
\mathrm{ICF}(\mathrm{Se}) = \mathrm{Se}/ \mathrm{Se}^{2+} = (2.411 + 0.5146x^{12.29} - 2.417e^{-0.1723x})^{-1}, \label{eq7}
\end{equation}
\begin{displaymath}
x = \mathrm{Ar}^{2+}/\mathrm{Ar} \geq 0.0344;
\end{displaymath}
\begin{equation}
\eqnum{8}
\mathrm{ICF}(\mathrm{Se}) = \mathrm{Se}/ \mathrm{Se}^{3+} = (0.03238 + 0.3225x + 0.4013x^2)^{-1}, \label{eq8}
\end{equation}
\begin{displaymath}
x = \mathrm{O}^{2+}/\mathrm{O};
\end{displaymath}
and
\begin{equation}
\eqnum{9}
\mathrm{ICF}(\mathrm{Se}) = \mathrm{Se}/ (\mathrm{Se}^{2+} + \mathrm{Se}^{3+}) = (-0.3703 + 0.3558e^{1.27x})^{-1}, \label{eq9}
\end{equation}
\begin{displaymath}
x = (\mathrm{Cl}^{2+} + \mathrm{Cl}^{3+}) / \mathrm{Cl} \geq 0.0315.
\end{displaymath}
We retain the equation numbering of \citetalias{sterling15} to avoid confusion.  Below the lower limits to $x$ for Equations~\ref{eq7} and \ref{eq9}, the ICFs are negative and thus invalid.  Equation~\ref{eq9} is expected to be the most accurate ICF, as it accounts for both Se$^{2+}$ and Se$^{3+}$, which reduces the magnitude and importance of uncertainties in the correction for unobserved ions.

Most nebular Se abundances have been derived with Equation~\ref{eq8}, since Se$^{3+}$ has thus far been the only Se ion unambiguously detected in most PNe.  Using this ICF, \citetalias{sterling15} and \citet{mashburn16} found that some PNe that exhibit \emph{s}-process enhancements of Kr are not enriched in Se -- in some cases, the Se abundance relative to O or Ar is subsolar.  In Galactic PNe exhibiting both Se and Kr emission, $[$Kr/Se$]=0.5\pm$0.2 \citepalias{sterling15}, which is larger than the values of 0.1--0.2~dex predicted by recent AGB nucleosynthesis models \citep[e.g.,][]{cristallo15, karakas16}.  The discrepancy with models raises the question of whether Equation~\ref{eq8} underestimates elemental Se abundances, or if the difference can be attributed to observational uncertainties.

Values for the Se abundance in NGC~5315, derived with Equations~\ref{eq7}--\ref{eq9}, are given in Table~\ref{ionicf}.  Due to the uncertainty of the $[$\ion{Se}{3}$]$~0.8858~\micron\ flux after correcting for \ion{He}{1} contamination, we utilize only the 1.0994~\micron\ line for the Se$^{2+}$/H$^+$ abundance.  Interestingly, Equations~\ref{eq7} and \ref{eq9} give larger Se abundances in NGC~5315 than Equation~\ref{eq8}.  The Se abundance from Equation~\ref{eq9} combined with $[$Kr/H$]$ from \citet{madonna17} gives $[$Kr/Se$]=-0.10\pm$0.22.  In contrast, the Se abundance is subsolar if derived using Equation~\ref{eq8}, resulting in $[$Kr/Se$]=0.48\pm$0.15.

However, conclusions regarding the accuracy of the Se ICFs should not be drawn from a single PN.  The discrepancy in the Se abundances from different ICFs may be due to observational uncertainties, inaccuracies in the atomic data governing the ionization balance of Se, or a breakdown of Equation~8 in PNe with excitation levels similar to NGC~5315 \citepalias[as seen for the Kr ICF Equation~3 of][]{sterling15}.  Observations of $[$\ion{Se}{3}$]$~1.0994~\micron\ in additional PNe, with a range of excitation levels, are needed to fully test the Se ICFs of \citetalias{sterling15}.  

\subsection{Kr$^{5+}$ Abundances and New Kr Ionization Correction Factors} \label{kr6_sect}

The detection of $[$\ion{Kr}{6}$]$ enables more accurate nebular Kr abundance determinations, although the effect is not as large as that of $[$\ion{Se}{3}$]$ on Se abundances due to the relatively small Kr$^{5+}$ fractional abundance.  In the PNe we consider, the Kr$^{5+}$ abundance is 2--20 times smaller than Kr$^{2+}$ and/or Kr$^{3+}$ \citep{sharpee07, garcia-rojas15, mashburn16}, the two most dominant Kr ions in PNe. 

\begin{figure}[t!]
\epsscale{0.5}
\plotone{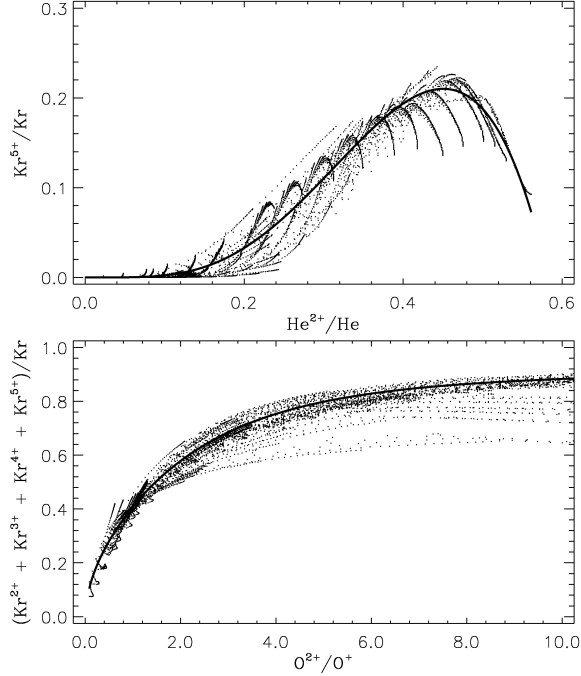}
\caption{Correlations between the fractional abundances of Kr$^{5+}$ and He$^{2+}$ (top panel) and the sum of detected Kr ions and the O$^{2+}$/O$^+$ ratio, from \citetalias{sterling15}'s grid of Cloudy models.  The correlations shown are for the Cloudy ``default'' PN abundances (C-rich) without dust grains, with each dot representing a single model out of 10,471 in the grid.  Similar correlations are found for other grids.  The thick solid lines represent the analytical fits to the correlations given in Equations~\ref{icf_kr6} and \ref{icf_kr3456}.}
\label{kr_icfs}
\end{figure}

To compute elemental Kr abundances from the $[$\ion{Kr}{6}$]$~1.2330~\micron\ line, we searched for correlations between Kr$^{5+}$ ionic fractions (and combinations of Kr ions including Kr$^{5+}$) and those of commonly-detected lighter species in the grids of Cloudy \citep{ferland13} models of \citetalias{sterling15}.  The strongest correlations are depicted in Figure~\ref{kr_icfs}.  The Kr$^{5+}$ fractional abundance is correlated with that of He$^{2+}$, albeit with a relatively large dispersion that increases for grids that include dust or have subsolar metallicities.  No other significant correlations were found for Kr$^{5+}$, and thus the ICF to convert Kr$^{5+}$/H$^+$ to elemental Kr/H abundances is uncertain.  In contrast, the fraction of all detected Kr ions in PNe (Kr$^{2+}$--Kr$^{5+}$) shows a tighter correlation with the O$^{2+}$/O$^+$ ratio.  

We fit each correlation with an analytic function, the inverse of which serves as an ICF:
\begin{equation}
\eqnum{1}
\mathrm{ICF}(\mathrm{Kr}) = \mathrm{Kr}/ \mathrm{Kr}^{5+} = (2.881\times10^{-5} - 16.58x^{7.215} + 474.1x^{4.974}e^{-7.84x})^{-1}, \label{icf_kr6}
\end{equation}
\begin{displaymath}
x = \mathrm{He}^{2+}/\mathrm{He};
\end{displaymath}
and
\begin{equation}
\eqnum{2}
\mathrm{ICF}(\mathrm{Kr}) = \mathrm{Kr}/ (\mathrm{Kr}^{2+} + \mathrm{Kr}^{3+} + \mathrm{Kr}^{4+} + \mathrm{Kr}^{5+}) = (-1206.83 + 1207.62x^{3.568\times10^{-5}} - 0.6035e^{-0.4583x})^{-1}, \label{icf_kr3456}
\end{equation}
\begin{displaymath}
x = \mathrm{O}^{2+}/\mathrm{O}^+ \geq 0.01671.
\end{displaymath}
In Equation~\ref{icf_kr3456}, the ICF is negative for O$^{2+}$/O$^+$ values below the denoted limit.

Equation~\ref{icf_kr3456}, which primarily corrects for Kr$^+$, produces Kr abundances that agree well with results from the literature (Table~\ref{ionicf}).  Equation~\ref{icf_kr6} tends to underestimate the Kr abundance compared to other ICFs, and shows a possible trend with nebular excitation.  The Kr abundance derived with this equation agrees within the uncertainties with other estimates in SMP~99 \citep{mashburn16}, which has He$^{2+}$/He~=~0.19 \citep{ld06}, but is lower than previous estimates by factors of 4--10 for the other more highly excited PNe \citep[He$^{2+}$/He~$\approx0.3$-0.4;][]{zhang05, ld06, garcia-rojas15}.  The poor accuracy is perhaps unsurprising, given the small Kr$^{5+}$ ionic fractions and the uncertainties of the ICFs derived from Equation~\ref{icf_kr6}.  Observations of $[$\ion{Kr}{6}$]$ in additional PNe, paired with deep optical spectra in which other Kr ions are detected, are needed to fully test the accuracy of Equation~\ref{icf_kr6} and determine whether the discrepancies with other ICFs can be attributed to observational uncertainties or the ICF itself.

\section{CONCLUSIONS}

The detections of $[$\ion{Se}{3}$]$~1.0994 and $[$\ion{Kr}{6}$]$~1.2330~\micron\ provide the means to improve the accuracy of nebular Kr and especially Se abundances.  These are the most widely-detected \emph{n}-capture elements in astrophysical nebulae, and have been used to study \emph{s}-process enrichments in numerous PNe.  We compute collision strengths for each ion, and use these to derive ionic abundances.  We test the Se ICF prescriptions of \citetalias{sterling15}, finding a larger Se abundance in NGC~5315 with ICFs that include Se$^{2+}$/H$^+$ than the ICF that relies only on Se$^{3+}$/H$^+$.  We also derive Kr ICFs that incorporate Kr$^{5+}$/H$^+$ abundances, and apply these to four PNe.  Additional observations of these lines in PNe with a range of excitation levels are needed to more rigorously test the Se ICFs of \citetalias{sterling15} and our new Kr ICFs.

\acknowledgments

We thank A.\ Karakas for discussions of Kr/Se ratios produced by \emph{s}-process nucleosynthesis.  NCS acknowledges support from NSF award~AST-0901432, IUR from NSF grant PHY~14-30152 (Physics Frontier Center/JINA-CEE), and JG-R and SM from the Spanish Ministry of Economy and Competitiveness (MINECO) grant AYA2015-65205-P and Severo Ochoa Program SEV-2015-0548.  SM acknowledges a Ph.D.\ fellowship from Instituto de Astrof\'isica de Canarias, and a Universidad de~La~Laguna travel grant supporting a visit to the University of West Georgia, where part of this work was done.  This work has made use of NASA's Astrophysics Data System.

\bibliographystyle{apj}

\end{document}